\documentclass[11pt]{article}
\usepackage{amssymb,latexsym,amsmath,amsbsy}
\usepackage[dvips]{graphicx}
\usepackage{hyperref}
\newtheorem{theo}{Theorem}
\newtheorem{proposition}[theo]{Proposition}
\renewcommand{\atop}[2]{\genfrac{}{}{0pt}{}{#1}{#2}}
\def\nn{\nonumber}

\def\su{\mathfrak{su}}

\begin{document}
\begin{center}
	{\Large \bf
		A finite quantum oscillator model related to \\[1mm]
		special sets of Racah polynomials
	} \\[5mm]
	{\bf Roy Oste\footnote{E-mail: Roy.Oste@UGent.be}, Joris Van der Jeugt\footnote{E-mail: Joris.VanderJeugt@UGent.be}}\\[1mm]
	Department of Applied Mathematics, Computer Science and Statistics, Ghent University,\\
	Krijgslaan 281-S9, B-9000 Gent, Belgium.
\end{center}

\begin{abstract}
In~\cite{DoubleHahn} we classified all pairs of recurrence relations in which two (dual) Hahn polynomials with different parameters appear. Such pairs are referred to as (dual) Hahn doubles, and the same technique was then applied to obtain all Racah doubles. 
We now consider a special case concerning the doubles related to Racah polynomials.
This gives rise to an interesting class of two-diagonal matrices with closed form expressions for the eigenvalues.
Just as it was the case for (dual) Hahn doubles, the resulting two-diagonal matrix can be used to construct a finite oscillator model.
We discuss some properties of this oscillator model, give its (discrete) position wavefunctions explicitly, and illustrate their behaviour by means of some plots.
\end{abstract}

\setcounter{equation}{0}
\section{Introduction} \label{sec:Introduction}%

In a recent paper~\cite{DoubleHahn} all pairs of recurrence relations in which two Hahn, dual Hahn or Racah polynomials with different parameters appear were classified. We used the term (dual) Hahn doubles or Racah doubles for such pairs. 
They were shown to correspond to Christoffel-Geronimus pairs of (dual) Hahn or Racah polynomials~\cite{DoubleHahn}.

In the present paper, we shall consider a special case of a Racah double.
This special case is chosen in such a way that the related two-diagonal (Jacobi) matrix $M$ has a very simple spectrum. 
The eigenvectors of $M$ can then be written in terms of the corresponding Racah polynomials.

The main reason to study the special case considered here is because it is particularly interesting in the framework of finite oscillator models.
Finite oscillator models were introduced and investigated in a number of papers, see e.g.~\cite{Atak-Suslov, Atak2001, Atak2001b, Atak2005, JSV2011, JSV2011b}.
The standard example is the $\su(2)$ oscillator model~\cite{Atak-Suslov,Atak2001}.
In brief, this model is based on the $\su(2)$ algebra with basis elements $J_0=J_z$, $J_\pm=J_x\pm J_y$ satisfying
\begin{equation}
[J_0,J_\pm]=\pm J_\pm, \qquad [J_+,J_-]=2J_0,
\end{equation}
with unitary representations of dimension $2j+1$ (where $j$ is integer or half-integer). 
Recall that the oscillator Lie algebra can be considered as an associative algebra (with unit element $1$) with three generators $\hat H$, $\hat q$ and $\hat p$ (the Hamiltonian, the position and the momentum operator) subject to
\begin{equation}
[\hat H, \hat q] = -i \hat p, \qquad [\hat H,\hat p] = i \hat q, \qquad [\hat q, \hat p]=i,
\label{Hqp}
\end{equation}
in units with mass and frequency both equal to~1, and $\hbar=1$.
The first two are the Hamilton-Lie equations; the third the canonical commutation relation. 
The canonical commutation relation is not compatible with a finite-dimensional Hilbert space.
Following this, one speaks of a finite oscillator model if $\hat H$, $\hat q$ and $\hat p$ belong to some algebra such that the Hamilton-Lie equations are satisfied and such that the spectrum of $\hat H$ in representations of that algebra is equidistant~\cite{Atak2001,JSV2011}. 

In the $\su(2)$ model, one chooses
\begin{equation}
\hat H=J_0+j+\frac12,\quad \hat q=\frac12(J_++J_-),\quad \hat p=\frac{i}{2}(J_+-J_-).
\label{su2-Hpq}
\end{equation}
These indeed satisfy $[\hat H, \hat q] = -i \hat p$, $[\hat H,\hat p] = i \hat q$, and the spectrum of $\hat H$
is equidistant in the representation $(j)$ labeled by $j$ (and given by $n+\frac12$; $n=0,1,\ldots,2j$).
Clearly, for this model the position operator $\hat q=\frac12(J_++J_-)$ also has a finite spectrum in the representation $(j)$ given by 
$q\in \{-j,-j+1,\ldots,+j\}$. 
In terms of the standard $J_0$-eigenvectors $|j,m\rangle$, the eigenvectors of $\hat q$ can be written as
\begin{equation}
|j,q) = \sum_{m=-j}^j \Phi_{j+m}(q) |j,m\rangle.
\label{4}
\end{equation}
The coefficients $\Phi_{n}(q)$ are the position wavefunctions, and in this model~\cite{Atak-Suslov,Atak2001} they turn out to be (normalized) symmetric Krawtchouk polynomials, $\Phi_n(q) \sim  K_ n(j+q;\frac12,2j)$.
The shape of the these wavefunctions is reminiscent of those of the canonical oscillator:
under the limit $j\rightarrow \infty$ they coincide with the 
canonical wavefunctions in terms of Hermite polynomials.

In the present paper we develop a related but new finite oscillator model, following the ideas of~\cite{JSV2011} where a dual Hahn double was used to extend the $\su(2)$ model. The recent classification~\cite{DoubleHahn} of all (dual) Hahn doubles and Racah doubles opens the way to investigate such new models. The basic ingredient is a special Racah double from this classification, that is explained and analysed in Section~2. In Section~3 we study the related finite oscillator model, and in particular we focus on some properties of the discrete position wavefunctions.

\section{Racah polynomials and two Racah doubles}

Racah polynomials $R_n(\lambda(x);\alpha,\beta,\gamma,\delta)$ of degree $n$ ($n=0,1,\ldots,N$) in the variable $\lambda(x)=x(x+\gamma+\delta+1)$
are defined by~\cite{Koekoek,Suslov,Ismail}
\begin{equation}
\label{Racah}
R_n(\lambda(x);\alpha,\beta,\gamma,\delta)=  {\;}_4F_3 \left( \atop{-n,n+\alpha+\beta+1,-x,x+\gamma+\delta+1} 
{\alpha+1,\beta+\delta+1,\gamma+1} ; 1 \right),
\end{equation}
where one of the denominator parameters should be $-N$:
\begin{equation}
\label{-N}
\alpha+1=-N\qquad\hbox{or}\qquad \beta+\delta+1=-N \qquad\hbox{or}\qquad \gamma+1=-N.
\end{equation}
Herein, the function $_4F_3$ is the generalized hypergeometric series~\cite{Bailey,Slater}:
\begin{equation}
	{\;}_pF_q \left( \atop{a_1,\dots,a_p}{b_1,\dots,b_q} ; z \right) = \sum_{k=0}^{\infty} \frac{(a_1)_k\dots(a_p)_k}{(b_1)_k\dots(b_q)_k}\frac{z^k}{k!},
\label{defF}
\end{equation}
where we use  the common notation for  Pochhammer symbols~\cite{Bailey,Slater}
$(a)_k=a(a+1)\cdots(a+k-1)$ for $k=1,2,\ldots$ and $(a)_0=1$. 
Note that in~(\ref{Racah}),  the series is terminating  
because of the appearance of the negative integer $-n$ as a numerator parameter. 

Racah polynomials satisfy a (discrete) orthogonality relation (depending on the choice of which parameter relates to $-N$)~\cite{Koekoek,NIST}. For the choice $\alpha+1=-N$ 
we have
\begin{equation}
\sum_{x=0}^N w(x;\alpha, \beta,\gamma,\delta) R_n(\lambda(x);\alpha,\beta,\gamma,\delta) R_{n'}(\lambda(x);\alpha,\beta,\gamma,\delta) = h_n(\alpha,\beta,\gamma,\delta)\, \delta_{n,n'},
\label{orth-Q}
\end{equation} 
where
\begin{align}
& w(x;\alpha, \beta,\gamma,\delta) = \frac{(\alpha+1)_x(\beta+\delta+1)_x(\gamma+1)_x(\gamma+\delta+1)_x((\gamma+\delta+3)/2)_x}{(-\alpha+\gamma+\delta+1)_x(-\beta+\gamma+1)_x	((\gamma+\delta+1)/2)_x(\delta+1)_x x!}, \nn \\
& h_n(\alpha,\beta,\gamma,\delta)= \frac{(-\beta)_N(\gamma+\delta+2)_N}{(-\beta+\gamma+1)_N(\delta+1)_N } \label{wh} \\ 
&\qquad\times\frac{(n+\alpha+\beta+1)_n (\alpha+\beta-\gamma+1)_n (\alpha-\delta+1)_n(\beta+1)_n n!}{(\alpha+\beta+2)_{2n}	(\alpha+1)_n (\beta+\delta+1)_n (\gamma+1)_n}. \nn
\end{align}
Under certain restrictions such as $\gamma,\delta>-1$ and $\beta>N+\gamma$ or $\beta<-N-\delta-1$, which ensure positivity of the functions $w$ and $h$, we can define orthonormal Racah functions as follows:
\begin{equation}\label{R-tilde}
\tilde R_n(\lambda(x);\alpha, \beta,\gamma,\delta) \equiv \frac{\sqrt{w(x;\alpha,\beta,\gamma,\delta)}\, R_n(\lambda(x);\alpha, \beta,\gamma,\delta)}{\sqrt{h_n(\alpha,\beta,\gamma,\delta)}}.
\end{equation}

After settling this notation, let us turn to a result from~\cite{DoubleHahn}. 
The matrices appearing here will always be of a special tridiagonal form, namely
\begin{equation}
\label{MK}
M= \left( \begin{array}{ccccc}
             0 & M_0  &    0   &   \phantom{\ddots}     &      \\
            M_0 &  0  &  M_1   &  0  &      \\
              0  & M_1  &   0  & M_2 &  \ddots   \\
             \phantom{\ddots}    & 0 & M_2 & 0 & \ddots \\
                 &       &    \ddots   &  \ddots  &  \ddots
          \end{array} \right),
\end{equation} 
and such matrices will be referred to as two-diagonal. 
The following two propositions were obtained in~\cite[Appendix]{DoubleHahn}:

\begin{proposition}
\label{prop1}
Let $\alpha+1=-N$, and suppose that $\gamma,\delta>-1$ and $\beta>N+\gamma$ or $\beta<-N-\delta-1$.
Consider two $(2N+2)\times(2N+2)$ matrices $U$ and $M$, defined as follows.
$U$ has elements ($n,x\in\{0,1,\ldots,N\}$):
\begin{align}
& U_{2n,N-x} = U_{2n,N+x+1} = \frac{(-1)^n}{\sqrt{2}} \tilde R_n(\lambda(x);\alpha,\beta,\gamma,\delta+1), \nn\\
& U_{2n+1,N-x} = -U_{2n+1,N+x+1} = -\frac{(-1)^n}{\sqrt{2}} \tilde R_n(\lambda(x);\alpha,\beta+1,\gamma+1,\delta); \label{R-I}
\end{align}
$M$ is the two-diagonal $(2N+2)\times(2N+2)$-matrix of the form~\eqref{MK} with
\begin{align}
M_{2k} &= 2\sqrt{ \frac{(N-\beta-k)(\gamma+1+k)(N+\delta+1-k)(k+\beta+1)	}{(N-\beta-2k)(2k-N+1+\beta) }},\nn\\
M_{2k+1} &= 2\sqrt{  \frac{(\gamma+N-\beta-k)(k+1)(N-k)(k+\delta+\beta+2) }{(N-\beta-2k-2)(2k-N+1+\beta) }}.
\label{M_k-RI}
\end{align}
Then $U$ is orthogonal, and the columns of $U$ are the eigenvectors of $M$, i.e.\ $M U = U D$,
where $D$ is a diagonal matrix containing the eigenvalues of $M$:
\begin{align}
& D= \mathop{\rm diag}\nolimits  (-\epsilon_N,\ldots,-\epsilon_1,-\epsilon_0,\epsilon_0,\epsilon_1,\ldots,\epsilon_{N}), \nn\\
& \epsilon_{k}=2\sqrt{(k+\gamma+1)(k+\delta+1)}\qquad (k=0,1,\ldots,N). 
\label{eigenv-even}
\end{align}
\end{proposition}
In short, the pair of polynomials $R_n(\lambda(x);\alpha,\beta,\gamma,\delta+1)$ and $R_n(\lambda(x);\alpha,\beta+1,\gamma+1,\delta)$ form a ``Racah double'', and the relation $MU=UD$ governs the corresponding recurrence relations with $M$ taking the role of a Jacobi matrix~\cite{DoubleHahn}.

\begin{proposition}
\label{prop2}
Let $\alpha+1=-N$, and suppose that $\gamma,\delta>-1$ and $\beta>N+\gamma$ or $\beta<-N-\delta$.
Consider two $(2N+1)\times(2N+1)$ matrices $U$ and $M$, defined as follows.
\begin{align}
& U_{2n,N-x} = U_{2n,N+x} = \frac{(-1)^n}{\sqrt{2}} \tilde R_n(\lambda(x);\alpha,\beta,\gamma,\delta), \qquad(n=0,\ldots,N;\ x=1,\ldots,N)  \nn \\
& U_{2n+1,N-x-1} = -U_{2n+1,N+x+1} = -\frac{(-1)^n}{\sqrt{2}} \tilde R_n(\lambda(x);\alpha+1,\beta,\gamma+1,\delta+1), \label{R-III}  \\
& \qquad \qquad \qquad(n,x\in\{0,\ldots,N-1\}) \nn \\
& U_{2n,N}= (-1)^n \tilde R_n(\lambda(0);\alpha,\beta,\gamma,\delta),\qquad U_{2n+1,N}=0. \nn
\end{align}
$M$ is the two-diagonal $(2N+1)\times(2N+1)$-matrix of the form~\eqref{MK} with
\begin{align}
M_{2k} &= 2\sqrt{ \frac{(\gamma+k+1)(-N+\beta+k)(N-k)(k+\delta+\beta+1)	}{(N-\beta-2k)(N-\beta-2k-1) }},\nn\\
M_{2k+1} &= 2\sqrt{  \frac{(\gamma+N-\beta-k)(k+1)(k+\beta+1)(k-\delta-N) }{(N-\beta-2k-2)(N-\beta-2k-1) }}.
\label{M_k-RIII}
\end{align}
Then $U$ is orthogonal, and the columns of $U$ are the eigenvectors of $M$, i.e.\ $M U = U D$,
where $D$ is a diagonal matrix containing the eigenvalues of $M$:
\begin{align}
& D= \mathop{\rm diag}\nolimits  (-\epsilon_N,\ldots,-\epsilon_1,0,\epsilon_1,\ldots,\epsilon_{N}), \nn\\
& \epsilon_{k}=2\sqrt{k(k+\gamma+\delta+1)}\qquad (k=1,\ldots,N). 
\label{eigenv-odd}
\end{align}
\end{proposition}

The special case considered in this paper is for $\gamma=\delta=-1/2$. The reason for this will be clear in the following, but at this point one can already observe that for these values the eigenvalues of $D$ (both in even dimensions, \eqref{eigenv-even}, as in odd dimensions, \eqref{eigenv-odd}) take a simple form.
For these special values, the matrix elements of $M$ in the case of Proposition~\ref{prop1} become:
\begin{align}
M_{2k} &= 2\sqrt{ \frac{(k-N+\beta)(k+1/2)(N-k+1/2)(k+\beta+1)	}{(2k-N+\beta)(2k-N+1+\beta) }},\nn\\
M_{2k+1} &= 2\sqrt{  \frac{(k-N+\beta+1/2)(k+1)(N-k)(k+\beta+3/2) }{(2k-N+\beta+2)(2k-N+1+\beta) }}. \nn
\end{align}
We see that in this case the expressions for coefficients with even and odd indices coincide and can be written as a single expression, namely
\begin{equation}
M_{k} = \sqrt{ \frac{(k+1)(2N+1-k)(k-2N+2\beta)(k+2\beta+2)	}{(2k-2N+2\beta)(2k-2N+2\beta+2) }},
\end{equation}
with $k\in \{0,\dots,2N\}$.
Suppose we are in the case $\beta>N+\gamma$, i.e.\ $\beta>N-1/2$. It will be useful to rewrite $2\beta=2N-1+c$, with $c>0$, and then the matrix elements take the form
\begin{equation}
M_{k}	 = \sqrt{ \frac{(k+1)(2N+1-k)(k -1 +c)(k+2N+1  +c)	}{(2k-1 +c)(2k+1 +c) }}, \qquad k\in\{0,\ldots, 2N\}.
\end{equation}

Also in the case of Proposition~\ref{prop2} the matrix elements of $M$ simplify for the special values $\gamma=\delta=-1/2$. They can also be written as a single expression, and after writing $2\beta=2N-1+c$ ($c>0$) they read:
\begin{equation}
	M_{k}	 = \sqrt{ \frac{(k+1)(2N-k)(k -1 +c)(k+2N  +c)	}{(2k-1 +c)(2k+1 +c) }}, \qquad k\in\{0,\ldots, 2N-1\}.
\end{equation}

Taking into account the size of the matrices in both cases, the results from Proposition~\ref{prop1} and Proposition~\ref{prop2} can be unified in the following:

\begin{proposition}
\label{propR1}
For $d$ a positive integer, $k\in \{0,\dots,d-1\}$ and a parameter $c>0$, let
\begin{equation}
	\label{Mkd}
	M_{k}	 =  \sqrt{ \frac{(k+1)(d-k)(k -1 +c)(k+d  +c)	}{(2k-1 +c)(2k+1 +c) }}.
\end{equation}
The eigenvalues of the tridiagonal $(d+1)\times(d+1)$-matrix of the form~\eqref{MK}	are given by the integers
\begin{equation}
	-d,-d+2,-d+4,\ldots,d-4,d-2,d
\label{KacEig}
\end{equation}
which are equidistant, symmetric around zero, and range from $-d$ to $d$. Hence for even $d=2N$, they are $d+1$ consecutive even integers, while for odd $d=2N+1$ they are $d+1$ consecutive odd integers. \\
For $d=2N$ even, the eigenvectors of $M$ are the columns of the matrix $U$ given by~\eqref{R-III}, with $\alpha=-N-1$, $\beta=N-1/2+c/2$, $\gamma=-1/2$ and $\delta=-1/2$.\\
For $d=2N+1$ odd, the eigenvectors of $M$ are the columns of the matrix $U$ given by~\eqref{R-I}, again with $\alpha=-N-1$, $\beta=N-1/2+c/2$, $\gamma=-1/2$ and $\delta=-1/2$.
\end{proposition}
Note in particular that the eigenvalues of $M$ are independent of the value of the parameter $c$, but of course $c$ appears in the expressions of the eigenvectors.

\section{A quantum oscillator model based on Racah polynomials} 

We now consider a one-dimensional quantum oscillator model based on the findings of the previous section. 
Particularly interesting about this model is that it contains a parameter $c>0$. 
By construction, the spectrum of the position operator in this model will be independent of the parameter $c$, equidistant and coincide with the spectrum of the $\su(2)$ finite oscillator model~\cite{Atak2001}. 

Let us first return to the $\su(2)$ model, briefly introduced in Section~1.
Working in a representation $(j)$ of dimension $2j+1$ (where $j$ is integer or half-integer), and in the standard basis $|j,m\rangle$ in which $J_0$ is diagonal, it follows from~\eqref{su2-Hpq} that the Hamiltonian is a diagonal matrix,
\begin{equation}
\hat H = \mathop{\rm diag}\nolimits  (\frac12,\frac32,\frac52,\dotsc,2j+\frac12) . 
\label{H1}
\end{equation}
In this context, it is more common to rewrite the basis vectors $|j,m\rangle$ ($m=-j,-j+1,\ldots,j$) of this representation space as $|n\rangle \equiv |j,n-j\rangle$ ($n=0,1,\ldots,2j$). Thus we can write
\begin{equation}
\hat H |n\rangle =  \left(n+\frac12\right)|n\rangle, \qquad (n=0,1,\ldots,2j).
\label{H2}
\end{equation}
Also from~\eqref{su2-Hpq}, the matrix form of the position operator $\hat q$ in this basis is given by
\begin{equation}
\hat q = \frac12 \left( \begin{array}{ccccc}
             0 & \mu_0  &    0   &   \phantom{\ddots}     &      \\
            \mu_0 &  0  &  \mu_1   &  0  &      \\
              0  & \mu_1  &   0  & \mu_2 &  \ddots   \\
             \phantom{\ddots}    & 0 & \mu_2 & 0 & \ddots \\
                 &       &    \ddots   &  \ddots  &  \ddots
          \end{array} \right), \qquad
					\mu_k=\sqrt{(k+1)(2j-k)},
\label{hatq}
\end{equation}					
and the momentum operator $\hat p$ takes the form
\begin{equation}
\hat p = \frac{i}{2} \left( \begin{array}{ccccc}
             0 & -\mu_0  &    0   &   \phantom{\ddots}     &      \\
            \mu_0 &  0  &  -\mu_1   &  0  &      \\
              0  & \mu_1  &   0  & -\mu_2 &  \ddots   \\
             \phantom{\ddots}    & 0 & \mu_2 & 0 & \ddots \\
                 &       &    \ddots   &  \ddots  &  \ddots
          \end{array} \right), \qquad
					\mu_k=\sqrt{(k+1)(2j-k)}.
\label{hatp}
\end{equation}	
Clearly, these operators (and matrix representations) satisfy the Hamilton-Lie equations from~\eqref{Hqp} (but not the canonical commutation relation).

Let us now turn to a new finite oscillator model based on the Racah polynomials introduced in the previous section.
For this purpose, observe that for any dimension $d+1=2j+1$, there is a close relationship between the matrix elements of $M$, given by~\eqref{Mkd}, and those of the above matrix~\eqref{hatq}:
\begin{equation}
M_{k}	 =  \sqrt{ (k+1)(2j-k)\frac{(c+k-1)(c+k+2j)}{(c+2k-1)(c+2k+1)}}, \qquad\qquad
\mu_k=\sqrt{(k+1)(2j-k)}.
\label{M-mu}
\end{equation}
Indeed, the positive parameter $c$ appearing in $M_k$ can be seen as a ``deformation'' of the element $\mu_k$. And clearly, in the limit $c\rightarrow +\infty$ one has that $M_k \rightarrow \mu_k$.
Following this, the elements of the new finite oscillator model -- in any dimension $2j+1$ ($j$ integer or half-integer) -- are defined as follows: the Hamiltonian $\hat H$ is the same operator as in~\eqref{H1} or~\eqref{H2}; the operators $\hat q$ and $\hat p$ are 
\begin{equation}
\hat q =\frac12 M = \frac12 \left( \begin{array}{ccccc}
             0 & M_0  &    0   &   \phantom{\ddots}     &      \\
            M_0 &  0  &  M_1   &  0  &      \\
              0  & M_1  &   0  & M_2 &  \ddots   \\
             \phantom{\ddots}    & 0 & M_2 & 0 & \ddots \\
                 &       &    \ddots   &  \ddots  &  \ddots
          \end{array} \right), \qquad
\hat p = \frac{i}{2} \left( \begin{array}{ccccc}
             0 & -M_0  &    0   &   \phantom{\ddots}     &      \\
            M_0 &  0  &  -M_1   &  0  &      \\
              0  & M_1  &   0  & -M_2 &  \ddots   \\
             \phantom{\ddots}    & 0 & M_2 & 0 & \ddots \\
                 &       &    \ddots   &  \ddots  &  \ddots
          \end{array} \right),					
\label{newqp}
\end{equation}
with $M_k$ given by~\eqref{Mkd} or equivalently~\eqref{M-mu}.

For this new model, the Hamiltonian-Lie equations are satisfied. 
So let us turn our attention to the properties of the position operator $\hat q$ (the properties of the momentum operator $\hat p$ are completely similar and will not be given explicitly).
Following Proposition~\ref{propR1}, the spectrum of $\hat q = \frac12 M$ is simply given by
\begin{equation}
-j, -j+1, -j+2, \ldots, j-2, j-1, j. 
\label{q-spectrum}
\end{equation}
Quite surprisingly, this spectrum is independent of the parameter $c$ appearing in the matrix elements~\eqref{newqp} of $\hat q$; but of course this is a consequence of Proposition~\ref{propR1}, and in particular of the special choice of $\gamma$ and $\delta$ earlier on in Section~2.
So the spectrum of $\hat q$ in the new model is just the same as in the familiar $\su(2)$ model.
For the eigenvectors of $\hat q$, however, things are different, as follows from the last part of Proposition~\ref{propR1}.
The orthonormal eigenvector of the position operator $\hat q$ for the eigenvalue $q$, denoted by $|q)$, is given in terms of the eigenstate basis of $\hat H$ by
\begin{equation}
|q) = \sum_{n=0}^{2j} U_{n,j+q} |n\rangle, \qquad q\in\{-j,-j+1 \ldots,j-1,j\}.
\label{qn}
\end{equation}
Herein, $U=(U_{kl})_{0\leq k,l \leq 2j}$ is the $(2j+1)\times(2j+1)$ matrix with elements defined in terms of normalized Racah polynomials~\eqref{R-tilde} as in the previous section.
Explicitly, for $j$ a half-integer, the elements of $U$ follow from Proposition~\ref{prop1}, with $N=j-1/2$ and $n,x\in\{0,\dotsc,N\}$:
\begin{align}
& U_{2n,N-x} = U_{2n,N+x+1} = \frac{(-1)^n}{\sqrt{2}} \tilde R_n(\lambda(x);-N-1,N-1/2+c/2,-1/2,1/2), \nn\\
& U_{2n+1,N-x} = -U_{2n+1,N+x+1} = -\frac{(-1)^n}{\sqrt{2}} \tilde R_n(\lambda(x);-N-1,N+1/2+c/2,1/2,-1/2).
\label{Ueven}
\end{align}
For $j$ an integer, they follow from Proposition~\ref{prop2}, with $N=j$:
\begin{align}
& U_{2n,j-x} = U_{2n,j+x} = \frac{(-1)^n}{\sqrt{2}} \tilde R_n(\lambda(x);-j-1,j-1/2+c/2,-1/2,-1/2),\nn\\ 
&\qquad\qquad\qquad(n=0,\ldots,j;\ x=1,\ldots,j)  \nn \\
& U_{2n+1,j-x-1} = -U_{2n+1,j+x+1} = -\frac{(-1)^n}{\sqrt{2}} \tilde R_n(\lambda(x);-j,j-1/2+c/2,1/2,1/2),\label{Uodd} \\ 
&\qquad\qquad\qquad(n,x\in\{0,\ldots,j-1\}) \nn \\
& U_{2n,j}= (-1)^n \tilde R_n(\lambda(0);-j-1,j-1/2+c/2,-1/2,-1/2),\qquad U_{2n+1,j}=0. \nn
\end{align}

These expressions deserve further attention.
Remember that, just as in~\eqref{4} for the $\su(2)$ model, quite generally 
the position (resp.\ momentum) wavefunctions are the overlaps between the normalized eigenstates of the position operator $\hat q$ (resp.\ the momentum operator $\hat p$) and the eigenstates of the Hamiltonian.
Let us denote the position wavefunctions for the new oscillator model by $\Phi^{(c)}_{n}(q)$, in order to emphasize the dependence upon the positive parameter~$c$.
We can thus write:
\begin{equation}
\Phi^{(c)}_{n}(q)  =   \langle n | q ) = U_{n,j+q}, 
\end{equation}
where $n=0,1,\ldots,2j$ and $q=-j,-j+1,\ldots,j-1,j$. So $\Phi^{(c)}_{0}(q)$ is the ``ground state'', $\Phi^{(c)}_{1}(q)$ the first excited state, and so on. 
All these expressions are real, and since we are dealing with a finite oscillator model they satisfy a discrete orthogonality relation:
\begin{equation}
\sum_{q=-j}^j \Phi^{(c)}_{n}(q) \Phi^{(c)}_{n'}(q) = \delta_{n,n'},\qquad
\sum_{n=0}^{2j} \Phi^{(c)}_{n}(q) \Phi^{(c)}_{n}(q') = \delta_{q,q'}.
\label{orth}
\end{equation}

Let us examine the explicit form of these functions in more detail, for the case $j$ half-integer
(the case $j$ integer is similar, and will not be treated explicitly). The expressions follow essentially from~\eqref{Ueven}.
The even wavefunctions are given by
\begin{equation}
	\Phi^{(c)}_{2n} (q)= \frac{(-1)^n}{\sqrt{2}} 
	\sqrt{W(n,  q; c ,j)} \
 {\;}_4F_3 \left( \atop{-q+1/2,q+1/2,-n,n+(c-1)/2} 
 {1/2,j+(c+1)/2,-j+1/2} ; 1 \right),
	\label{Phi-even}
\end{equation}
where
\[
W(n, q; c ,j) = \frac{w(|q|-1/2;-j-1/2,j-1+c/2,-1/2,1/2)}{h_n(-j-1/2,j-1+c/2,-1/2,1/2)}
\]
is written in terms of the weight function and square norm~\eqref{wh} of the Racah polynomials.
Note that $\Phi_{2n}^{(c)}(q)$ is indeed an even function of the position $q$, and depends on $q^2$ only.\\
The odd wavefunctions are given by
\begin{align}
	\Phi^{(c)}_{2n+1} (q) & = (-1)^n 
	 \sqrt{\frac{(4n+c+1)(2n+c-1)(2n+1)}{(4n+c-1)(2n+c+2j)(j-n)}} \nn \\ 
	& \times \sqrt{W(n,  q; c ,j)} {\;} \cdot q \cdot {\;}_4F_3 \left( \atop{-q+1/2,q+1/2,-n,n+(c+1)/2} 
	{3/2,j+(c+1)/2,-j+1/2} ; 1 \right).
	\label{Phi-odd}
\end{align}
Clearly, because of the factor $q$, $\Phi^{(c)}_{2n+1}(q)$ is an odd function of $q$.
The overall factor in~\eqref{Phi-odd}, by the way, arises from
\begin{align*}
&\frac{w(|q|-1/2;-j-1/2,j+c/2,1/2,-1/2)}{h_n(-j-1/2,j+c/2,1/2,-1/2)}=  \frac{	2 q^2 (4n+c+1)(2n+c-1)(2n+1) }{ (4n+c-1)(2n+c+2j)(j-n)	} \\
& \qquad \times \frac{w(|q|-1/2;-j-1/2,j-1+c/2,-1/2,1/2)}{h_n(-j-1/2,j-1+c/2,-1/2,1/2)}.
\end{align*}

It is interesting to study these discrete wavefunctions for varying values of $c$. 
We know already that in the limit $c\rightarrow +\infty$ the position operator ${\hat q}$ tends to the position operator of the $\su(2)$ model, so also the wavefunctions should have this behavior.
When $c$ tends to infinity, the wavefunctions $\Phi^{(c)}_n(q)$ are indeed Krawtchouk functions. 
Clearly, the ${}_4F_3$ series in~\eqref{Phi-even} and~\eqref{Phi-odd} reduce to ${}_3F_2$ series, which in turn reduce to ${}_2F_1$ series
according to
\begin{align}
& {\;}_3F_2 \left( \atop{-q+1/2,q+1/2,-n}{1/2,-j+1/2};1\right)
 = (-1)^n \frac{\binom{2j}{2n}}{\binom{j-1/2}{n}} 
 {\;}_2F_1 \left( \atop{-2n,-j-q}{-2j};2\right),\\
& {\;}_3F_2 \left( \atop{-q+1/2,q+1/2,-n}{3/2,-j+1/2};1\right) 
= -\frac{(-1)^n}{2q} \frac{\binom{2j}{2n+1}}{\binom{j-1/2}{n}} 
 {\;}_2F_1 \left( \atop{-2n-1,-j-q}{-2j};2\right)
\rlap{\,.}
\end{align}
These reductions have been given in~\cite{SV2011} and can be obtained, e.g., from~\cite[(48)]{Atak2005}. 
The ${}_2F_1$ series in the right hand side correspond to symmetric Krawtchouk polynomials
(i.e.\ Krawtchouk polynomials with $p=1/2$~\cite{Koekoek}). 
When $j$ tends to infinity, they yield the ordinary oscillator wavefunctions~\cite{Atak2005}. 

For other values of $c$, let us examine some plots of the discrete wavefunctions.
In Figure~1, we give the plots of $\Phi^{(c)}_n(q)$ for $n=0$, $n=1$ and $n=2$, and for some fixed $j$-value $j=33/2$.
The purpose is to observe the behavior of the wavefunctions as the positive parameter $c$ varies.
With this in mind, we have plotted these functions for the following $c$-values:
\[
c=10^{-6},\quad c=0.5,\quad c=1.5,\quad c=2,\quad c=4,\quad c=8, \quad c=32.
\]
For large values of $c$, the discrete wavefunctions take indeed the shape of those of the $\su(2)$ model (which, in turn, tend to the canonical oscillator wavefunctions when $j$ tends to infinity).
The case $c=0$ is ruled out, but we have examined a $c$-value close to 0, for which the behavior is somewhat `degenerate'.
To our surprise, the value $c=2$ is a kind of transition value for the ground state. 
Just looking at the ground state ($n=0$), one observes that for $c<2$ the shape is like a cup, whereas for $c>2$ it is like a cap.
In order to explain this transition value, recall from~\eqref{Phi-even} that
\begin{equation}
\Phi^{(c)}_{0} (q)= \frac{1}{\sqrt{2}}\sqrt{W(0,  q; c ,j)} = \frac{1}{\sqrt{2}} 
\left(\frac{w(|q|-1/2;-j-1/2,j-1+c/2,-1/2,1/2)}{h_0(-j-1/2,j-1+c/2,-1/2,1/2)}\right)^{1/2}.
\end{equation}
Using~\eqref{wh},
\begin{equation}
w(|q|-1/2;-j-1/2,j-1+c/2,-1/2,1/2) = \frac{(-j+1/2)_{|q|-1/2} (j+c/2+1/2)_{|q|-1/2} }{(j+3/2)_{|q|-1/2} (-j-c/2+3/2)_{|q|-1/2}},
\end{equation}
and thus for $c=2$ one finds $w(|q|-1/2;-j-1/2,j-1+c/2,-1/2,1/2) = 1$. In other words, for this special transition value $c=2$, the ground state wavefunction $\Phi^{(c)}_{0} (q)$ is a constant function. 

To conclude, in the field of finite quantum oscillators the original $\su(2)$ model remains an interesting model because of two reasons: the simple equidistant spectrum of the position (and momentum) operator, and the behavior of the position wavefunctions (which really look like discrete versions of the canonical oscillator wavefunctions, and tend to them when $j$ is sufficiently large).
The new model introduced in this paper deforms the $\su(2)$ model by a parameter $c>0$. The spectrum of the position (and momentum) operator is the same and thus remains simple and equidistant. The wavefunctions are deformed by the parameter $c$, and tend to those of the $\su(2)$ model when $c$ goes to infinity. The wavefunctions themselves are written in terms of Racah polynomials, and originate from a Racah double~\cite{DoubleHahn}. The shape of the wavefunctions could open applications beyond those of the $\su(2)$ model.


\newpage

\begin{figure}[htb]
\includegraphics[scale=1]{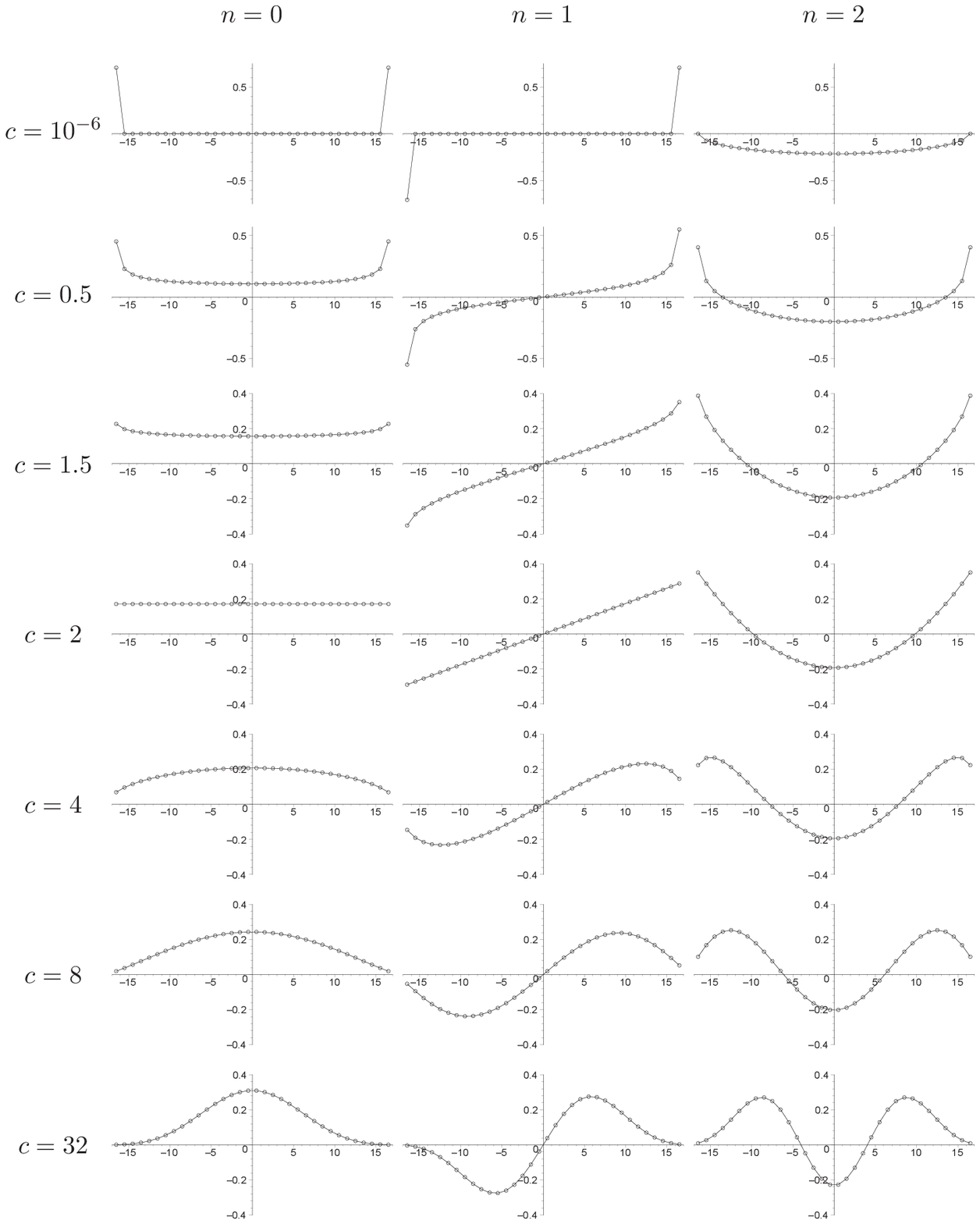} 
\caption{Plots of the discrete wavefunctions $\Phi^{(c)}_n(q)$ in the representation with $j=33/2$, for $n=0$ (left column), for $n=1$ (middle column) and for $n=2$ (right column). 
The wavefunctions are plotted for the following values of $c$ (from top to bottom): $10^{-6}, 0.5, 1.5, 2, 4, 8,32$.
}
\label{fig1}
\end{figure}

\end{document}